\documentclass[]{aastex631}

\graphicspath{{./}{figures/}}
\usepackage{easyReview}
\usepackage{comment}

\setreviewsoff

\begin{document}

\title{Origins of Very Low Helium Abundance Streams Detected in the Solar Wind Plasma}

\author[0000-0001-6018-9018]{Yogesh}
\affiliation{previously at Physical Research Laboratory, Navrangpura, Ahmedabad 380009, India}
\affiliation{NASA Goddard Space Flight Center, Greenbelt, MD, 20771, USA}
\affiliation{The Catholic University of America, Washington, DC 20064, USA}

\author{N. Gopalswamy}
\affiliation{NASA Goddard Space Flight Center, Greenbelt, MD, 20771, USA}

\author{D. Chakrabarty}
\affiliation{Physical Research Laboratory, Navrangpura, Ahmedabad 380009, India}

\author{Parisa Mostafavi}
\affiliation{Johns Hopkins University, Applied Physics Lab. Laurel, MD 20723, USA}

\author{Seiji Yashiro}
\affiliation{NASA Goddard Space Flight Center, Greenbelt, MD, 20771, USA}
\affiliation{The Catholic University of America, Washington, DC 20064, USA}

\author{Nandita Srivastava}
\affiliation{Udaipur Solar Observatory, Physical Research Laboratory, Udaipur, 313001, India}

\author[0000-0003-0602-6693]{Leon Ofman}
\affiliation{NASA Goddard Space Flight Center, Greenbelt, MD, 20771, USA}
\affiliation{The Catholic University of America, Washington, DC 20064, USA}
\affiliation{Visiting, Tel Aviv University, Tel Aviv, Israel}

\begin{abstract}

The abundance of helium ($A_{He}$) in the solar wind exhibits variations typically in the range from 2-5\% with respect to solar cycle activity and solar wind velocity. However, there are instances where the observed $A_{He}$ is exceptionally low ($<$ 1\%). These low-$A_{He}$ occurrences are detected both near the Sun and at 1 AU. The low $A_{He}$ events are generally observed near the heliospheric current sheet. We analyzed 28 low-$A_{He}$ events observed by the \textit{Wind} spacecraft and 4 by  Parker Solar Probe (PSP) to understand their origin. In this work, we make use of the ADAPT-WSA model to derive the sources of our events at the base of the solar corona. The modeling suggests that the low-$A_{He}$ events originated from the boundaries of coronal holes, primarily from large quiescent helmet streamers. We argue that the cusp above the core of the streamer can produce such very low helium abundance events. The streamer core serves as an ideal location for gravitational settling to occur as demonstrated by previous models, leading to the release of this plasma through reconnection near the cusp, resulting in low $A_{He}$ events. Furthermore, observations from Ulysses provide direct evidence that these events originated from coronal streamers.

\end{abstract}


\section{Introduction} \label{sec:intro}
Helium abundance ($A_{He}$) in the solar wind relative to protons is defined as $A_{He}$=100$\times N_{a}/N_p$, where $N_a$ and $N_p$ are the number densities of alpha particles (refereed interchangeably as `helium' in this paper, since He$^{++}$ is the dominant helium ion in the corona and in the solar wind) and protons. $A_{He}$ is known to vary significantly throughout the solar atmosphere \citep{Moses2020}. Depending on the sources, coronal, and interplanetary modulations, $A_{He}$ can range from 0.1\% to more than 30\% in number density ratio. If we consider the mass of He, it can account for the bulk of the solar wind mass flux at the high end of the abundance. $A_{He}$ follows the solar cycle (SC) and varies with solar wind velocity \citep{Kasper2007, Alterman2019, Yogesh2021, Alterman2021}. It has been shown that $A_{He}$ is 8\% in the photosphere and decreases to 4-5\% in the solar corona. It can increase up to 30\% in coronal mass ejections \citep[][and references therein]{Yogesh2022}. The enhancement of elemental abundances with low First Ionization Potential (FIP) in chromosphere, transition region, and coronal loops is known as the FIP effect. The FIP effect can also affect the relative abundances of heavy ions in the various coronal regions at the sources of the solar wind \citep[see the review by][]{Laming2015}. The FIP effect can enhance low ($<$10eV) FIP elements (e.g., Mg, Fe etc.) and reduce high FIP ($>$10eV) elements (e.g. He, Ne etc.) in the corona \citep{Laming2019}. The variation of the helium abundance relative to protons in coronal streamers was modeled with 2.5D three-fluid models, demonstrating the gravitational settling in the core of streamers and their expected observational signatures   \citep{Ofm04, Gio07, OfmKra2010, Abb16, Ofman2024}. The helium abundance also varies in the interplanetary (IP) medium depending on the interaction between different solar wind streams \citep{Durovcova2019, Yogesh2023}. Interactions between the helium ions and protons in the fast wind stream produced by Alfv\'{e}n waves in coronal holes were modelled \citep{Ofm04b}. The signatures of the periodic reconnections of the flux tubes carrying Alfv\'{e}n waves are also found in the abundances at 1AU \citep{Gershkovich2023, Kepko2024}. The variation in the helium abundance can tell us about the different processes occurring near the solar surface and interplanetary medium and the sources of the solar wind. The variation of $A_{He}$ in the solar wind \citep{Kasper2007, Alterman2019, Yogesh2021}, CMEs \citep{Fu2018, Yogesh2022}, SIRs \citep[][and references therein]{Yogesh2023}, has been explored by various researchers. However, there are very few works on the very low ($<$ 1\%) helium abundances in the solar wind. This is a significant problem as understanding very low helium abundances may provide important insights on the generation and propagation of the solar wind.   

\cite{Borrini1981} demonstrated on the basis of observations of IMP 6, 7, and 8 that the helium abundance is low near sector boundaries in the interplanetary (IP) medium. They also showed that these regions are generally associated with higher proton and electron densities, IP field polarity reversals, low bulk velocity, low alpha ($T_a$) and proton ($T_p$) temperature, minimum in the $T_a/T_p$ ratio and nearly equal proton and alpha speed. They suggested that these events may be related to the streamer belts and might be associated with the solar current sheets. However, the processes that could decrease $A_{He}$ so drastically were not addressed. Also, it remains unclear whether these reductions in the $A_{He}$ are because of the interplanetary modulations or due to processes closer to the Sun.  
 
Recently, a few other researchers \citep[e.g.,][]{Sanchez-Diaz2016, Vasquez2017} studied the low helium abundances using Helios spacecraft observations. A low helium abundance was detected in the very slow solar winds (VSSW, velocity $<$ 300km/s) \citep{Sanchez-Diaz2016}. They showed that $A_{He}$ in VSSW varies with the SC and the velocity of helium ions was less than that of protons in VSSW events. During solar maxima, $A_{He}$ value in the VSSW was higher than that in the normal slow wind (velocity $>$ 300km/s) in a few events. Although the main objective of these authors was to understand the properties of VSSW, they showed that these events were related to the Heliospheric Current Sheets (HCS) and High-Density Regions (HDR). \cite{Vasquez2017} studied $A_{He}$ in very slow ejecta and winds near the solar minima of SC23. They found that slow ejecta and winds show similar $A_{He}$ variation. Additionally, they observed that these events conform to the relationship between $A_{He}$ and solar wind velocity previously established by \cite{Kasper2007}. However, this relationship was not followed at a very low solar wind speed. The physical reason behind the low $A_{He}$ events and their sources have not been explored in depth so far.   

\cite{Woolley2021} and \citet{Ofm23} reported very low helium abundance ($<$1\%) observed by Parker Solar Probe (PSP) at perihelia. Based on data from the Ulysses and ACE spacecraft, \cite{Suess2009} found a significant reduction in helium abundance near quiescent current sheets. They suggested that the low $A_{He}$ events are generally observed near the HCS and streamers. The observation of low $A_{He}$ has been reported in the past by various researchers, but the process causing this depletion is unclear to a large extent. To address this gap, we critically examined the solar sources of these events. We have used the data from PSP, \textit{Wind}, ACE, and Ulysses to show that these events are observed through the heliosphere and originate from similar sources. We also make use of the Wang-Sheeley-Arge (WSA) model \citep{ArgePizzo2000,Arge2003a,Arge2004,McGregor2008} driven by Air Force Data Assimilative Photospheric Flux Transport (ADAPT: \citealp{Arge2009,Arge2010,Arge2013,Hickmann2015}) time-dependent photospheric field maps to derive the coronal magnetic field, as well as source regions of these events at 1 \(R_\odot\). The data used and model details are presented in Section \ref{sec:data}. Section \ref{sec:result} shows the results. The likely reasons behind the reduction in $A_{He}$ are discussed in Section 4.         

\section{Data selection and Model details} \label{sec:data}

We selected 28 events for which very low $A_{He}$ ($<$ 1\%) condition persists for more than 48 hours at the Sun-Earth L1 point using hourly averaged data from the Solar Wind Experiment \citep[SWE,][]{Ogilvie1995} onboard the \textit{Wind} spacecraft. Hourly averaged data removes the problems associated with transient spikes and data gaps in $A_{He}$. If we consider the full-cadence (92 seconds) data, $A_{He}$ remains below 1\% except for a few spikes. The data from SWE \citep{Ogilvie1995} and Magnetic Field Investigations (MFI) \citep{Lepping1995} instruments on board the \textit{Wind} satellite are used for proton, alpha parameters, and magnetic field, respectively. To identify the events, we used SWE data quality flag greater than 1, as our focus was on the alpha and proton densities. We find that most of the events are during the solar minima period,i.e.1995-1996, 2007-2010, 2017-2021. The list of the selected events can be found in supplementary table S1.

To analyze the low $A_{He}$ (four) events close to the Sun, we used data from the Solar Probe Analyzer for Ions (SPAN-I), which is a subsystem of the Solar Wind Electrons Alphas and Protons (SWEAP) \citep{Kasper2016} onboard PSP. SPAN-I has a Time-of-Flight section for mass-per-charge determination and can provide the alpha and proton parameters \citep{Livi2022}. The magnetic field observations are used from the FIELDS instrument suite \citep{Bale2016}. In this paper, we focused on times when the PSP made its closest approaches (i.e., `encounters'), ensuring that the spacecraft's tangential velocity was enough for SPAN-I to observe the core of the protons and alpha particle distributions. More details regarding the selection of appropriate data from SPAN-I can be found in \cite{Mostafavi2022}. We did not map the events between PSP and \textit{Wind}.

In this work, we use \textit{Wind} data to study the variation in solar wind parameters because of the availability of high cadence data. We also use the Advance Composition Explorer \citep[ACE; ][]{Stone1998} data to compare the ADAPT-WSA model output and interplanetary data. The \textit{Wind} and ACE observations have a high correlation in the case of most of the events. ACE’s Solar Wind Ion Composition Spectrometer \citep[SWICS;][]{Gloeckler1998} data are used to quantify compositional (Fe/O) changes during the events. We utilized SWICS 1.1 and SWICS 2.0 data (two-hour cadence) combined to span from 1998 to 2020. Data from the SWOOPS payload \citep{Bame1992} onboard Ulysses spacecraft are also used to provide the heliolatitude variation of helium abundance. 
	
We have used the ADAPT-WSA model to derive the coronal magnetic field and source region at 1 \(R_\odot\) for all the events. WSA is an empirical and physics-based model that derives the coronal field using a coupled set of potential field type models. The first is a traditional magnetostatic potential field source surface (PFSS) model \citep{Schatten1969, Altschuler1969, WangSheeley1992}, which determines the coronal field out to the source surface height. For this event, the traditional height of 2.5 \(R_\odot\) \citep{Hoeksema1983} is used. The PFSS solution then serves as input into the Schatten Current Sheet (SCS) model \citep{Schatten1971}, which provides a more realistic magnetic field topology of the upper corona (\textit{e.g.} from 2.5\,--\,21.5 \(R_\odot\)).  An empirical velocity relationship \citep{Arge2003b, Arge2004, Wallace2020} is then used to derive the solar wind speed at the outer coronal boundary, that is a function of both magnetic expansion factor as defined in \citet{WangSheeley1990} and the minimum angular separation between a field line foot point and the nearest coronal hole boundary (e.g., coronal hole boundary distance, as defined in \citealt{Riley2001, Riley2015}). The model then propagates solar wind parcels outward from the endpoints of each magnetic field line connected to the spacecraft while incorporating a simple 1-D modified kinematic model, which accounts for stream interactions by preventing fast streams from bypassing slow ones \citep{Arge2004}.  The model determines the time of arrival of these solar wind parcels at \textit{Wind}/ACE and PSP, allowing us to connect the in situ observed solar wind back to their model-determined solar origin. 

Synchronic photospheric field maps used as input to the WSA model were generated using the ADAPT model driven by Global Oscillations Network Group (GONG) magnetograms. ADAPT utilizes flux transport modelling \citep{WordenHarvey2000}, to account for solar time-dependent phenomena (\textit{e.g.} differential rotation, meridional and supergranulation flows) for locations on the Sun in which photospheric field measurements are not available (\textit{e.g. poles and far-side}). The best model output is determined by comparing the model-derived and observed radial magnetic field and solar wind speed.

\section{Results}\label{sec:result}

\subsection{In-situ measurements of low helium abundance events}

The low helium abundance ($A_{He}$ $<$1\%) events are identified using \textit{Wind} and PSP data. The events identified based on PSP data have shorter duration since the position of PSP changes rapidly at or near perihelia. PSP events are chosen based on SPAN-I data coverage, i.e., when alpha and proton measurements are both available and in the field of view of the instrument. Proper field-of-view criteria are important to avoid misinterpretation of PSP data. We found four such events from encounter 4 (2 events), 9 and 11 \citep[see,][for additional $A_{He}<1\%$ PSP events]{Ofm23}. The details regarding \textit{Wind} and PSP events are provided in the supplementary table S1.  

\begin{figure}
\begin{center}
\plotone{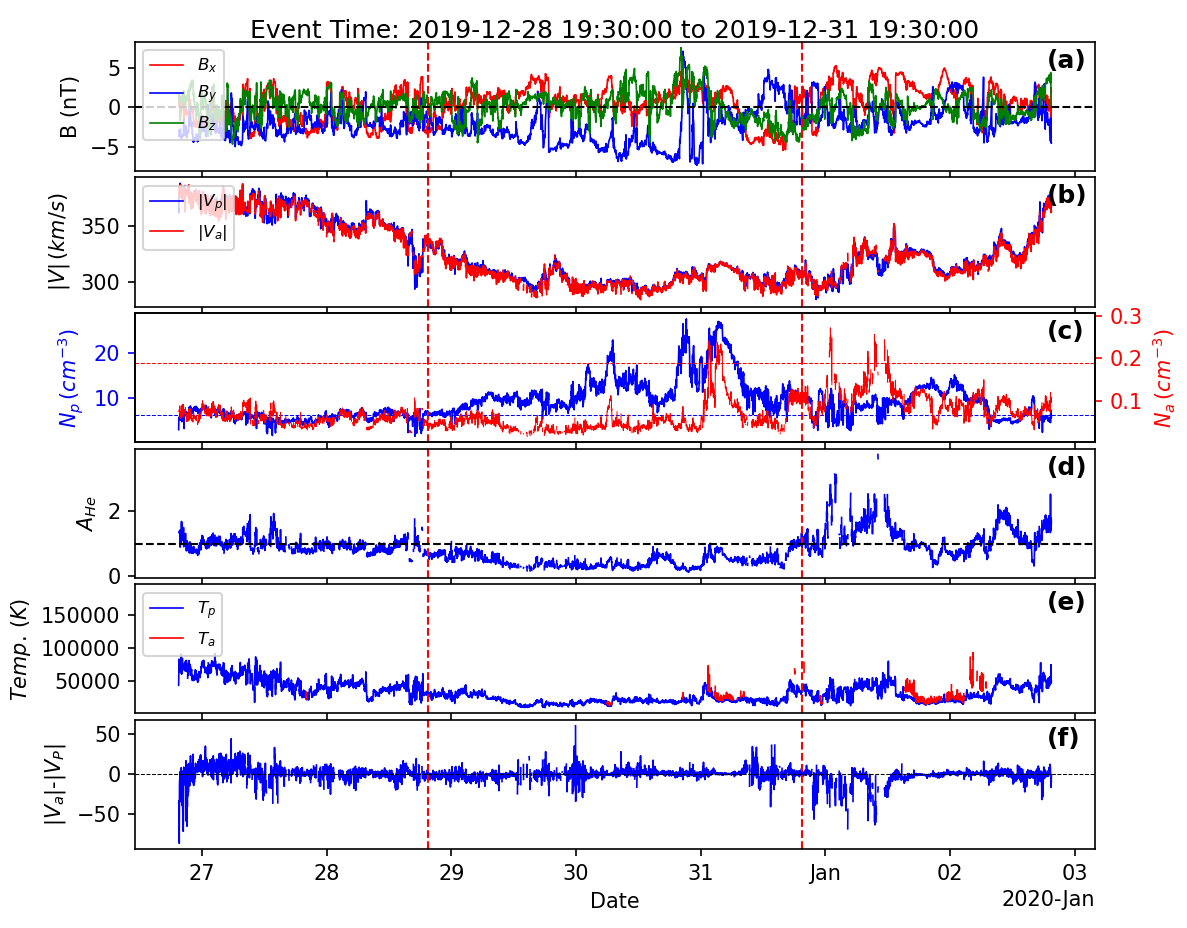}
\caption{A low helium abundance event observed by \textit{Wind} spacecraft in December 2019. The red vertical lines show the start and time of the event. The variations of the magnetic field components, speed (proton and alpha), number density (proton and alpha), helium abundance, proton and alpha temperature ($T_p$, $T_a$), and differential speed, are shown in panels $a$ to $f$. The blue and red horizontal lines in panel $c$ represent the average number density of protons and alpha particles over two solar cycles (SCs). Panel $c$ shows that the proton density ($N_p$) is higher than the average value of Np (6.29) over the two SCs. In panel c, the scale for proton density is on the left side and alpha density is on the right side. On the contrary, alpha particle densities are lower than the two solar cycle averaged values. There is significantly less speed difference between the alphas and protons (see panel $f$).         \label{fig:1}}

\end{center}
\end{figure}

\begin{figure}
\begin{center}
\plotone{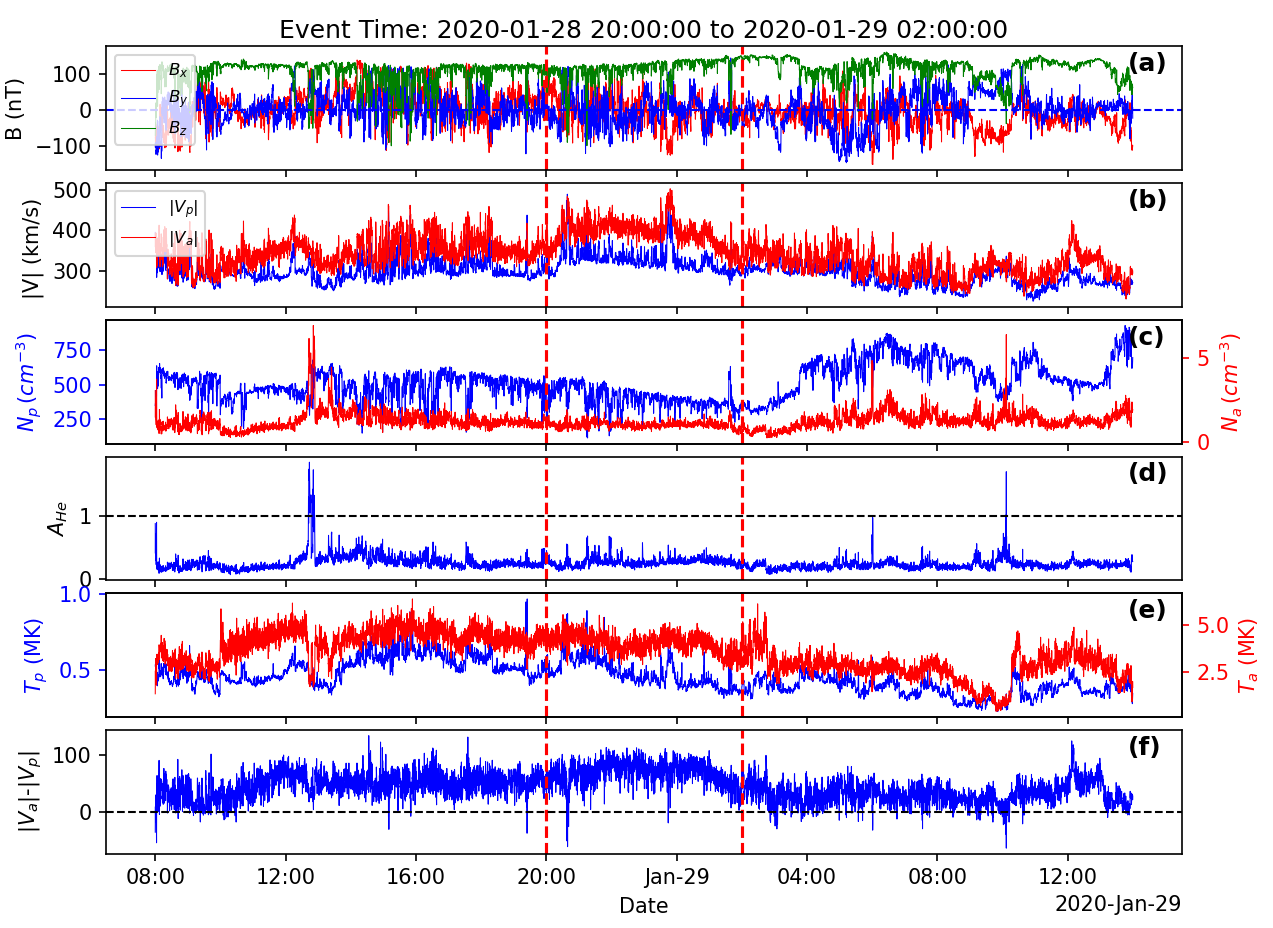}
\caption{The low helium abundance event observed by PSP in January 2020. Similar to Figure \ref{fig:1}, the variations in the solar wind parameters are shown in panels $a$ to $f$. The event time is selected based on the presence of alpha and protons in the field of view of the SPAN-I instrument onboard PSP. The vertical lines indicate the duration of the event during which the helium and hydrogen cores were within the field of view of the SPAN-I instrument and $A_{He}$ remained below 1\%. The dashed horizontal lines in panels a,d,f are B=0, $A_{He}$=1\% and zero differential speed, respectively.  \label{fig:2}}

\end{center}
\end{figure}

This section discusses the December 2019 event (number-21) observed by \textit{Wind} (see Fig. \ref{fig:1}). The other events are analyzed similarly, and the results are provided in the supplementary table S1. The Panels 1a-1f of Figure \ref{fig:1} present magnetic field components, speed (alpha and proton), number density (proton and alpha), helium abundance, the temperature of the proton ($T_p$) and alpha ($T_a$), and differential speed ($\triangle v=|v_a|-|v_p|$). It can be seen from Figure \ref{fig:1} that this marked event has low speed (average $\sim$310km/s). All the other events also show a similar lower speed. The blue and red horizontal dashed lines in Figure \ref{fig:1}c show the SC average of proton and alpha number density, respectively. It is seen that the proton number density is usually higher or equal to the average value almost all the time. In contrast, the helium number density is (with a few occasional spike-like increases) significantly less than the average helium number density during the interval marked by the vertical red dashed lines. The time of vertical lines is given in the plot title. Figure 1d plots $A_{He}$. The black dashed line in this panel is $A_{He}$=1\%. A similar parameter variation is observed for the other 28 events as well. This suggests that the reduction in alpha number density causes these very low $A_{He}$ events. We select intervals for which the helium abundance remains continuously below 1\% for more than 48 hours.

Figure \ref{fig:2} shows PSP event (number 2) from January 2020 from the supplementary Table S1. The event shown here is from encounter 4 when PSP was at a distance of 0.13 AU (27.8 $R_\odot$). The variation of all the parameters shown in Figure \ref{fig:2} is in the instrument frame. This does not impact our interpretation of the data because we are primarily focusing on density. The parameter format matches Figure 1.  Figure \ref{fig:2} (a-f) shows magnetic field components, the bulk velocity of alphas and protons, alpha and proton number density, helium abundance, $T_p$, $T_a$ and difference in the bulk velocities of alphas and protons. 

Figure \ref{fig:2} shows that the low $A_{He}$ interval (except for occasional spikes) is not only confined to the defined interval of encounter 4 but also extended more than 12 hours before and after the interval. Unlike in Figure \ref{fig:1}, there is a finite speed difference between the alphas and protons in the event duration. The differential speed close to the Sun is much larger than that at 1 AU. This is shown in \cite{Mostafavi2024}, who compared the observations at PSP and L1 data. The differential speed between the protons and alphas decreases as they propagate, possibly because of collisions they experience during propagation, or due to kinetic instabilities and wave-particle interaction \citep[e.g.,][]{Kasper2017, Alterman2018b, Durovcova2019}. Interestingly, this speed difference is reduced after the event, and the solar wind is also slower than average. This extended part also shows properties similar to the events observed at L1 by \textit{Wind}. The discrepancy between $T_a$ and $T_p$ observations at L1 and PSP are likely due to collisions and wave-particle interaction that the ions experience as they propagate \citep{Mostafavi2024}. Although this PSP event shows much lower $A_{He}$ as compared to \textit{Wind} events, other PSP events demonstrate $A_{He}$ values similar to those observed by the \textit{Wind} spacecraft.

\subsection{Back tracing of sources of low Helium abundance events}

\begin{figure}
\begin{center}
\includegraphics[scale=0.6]{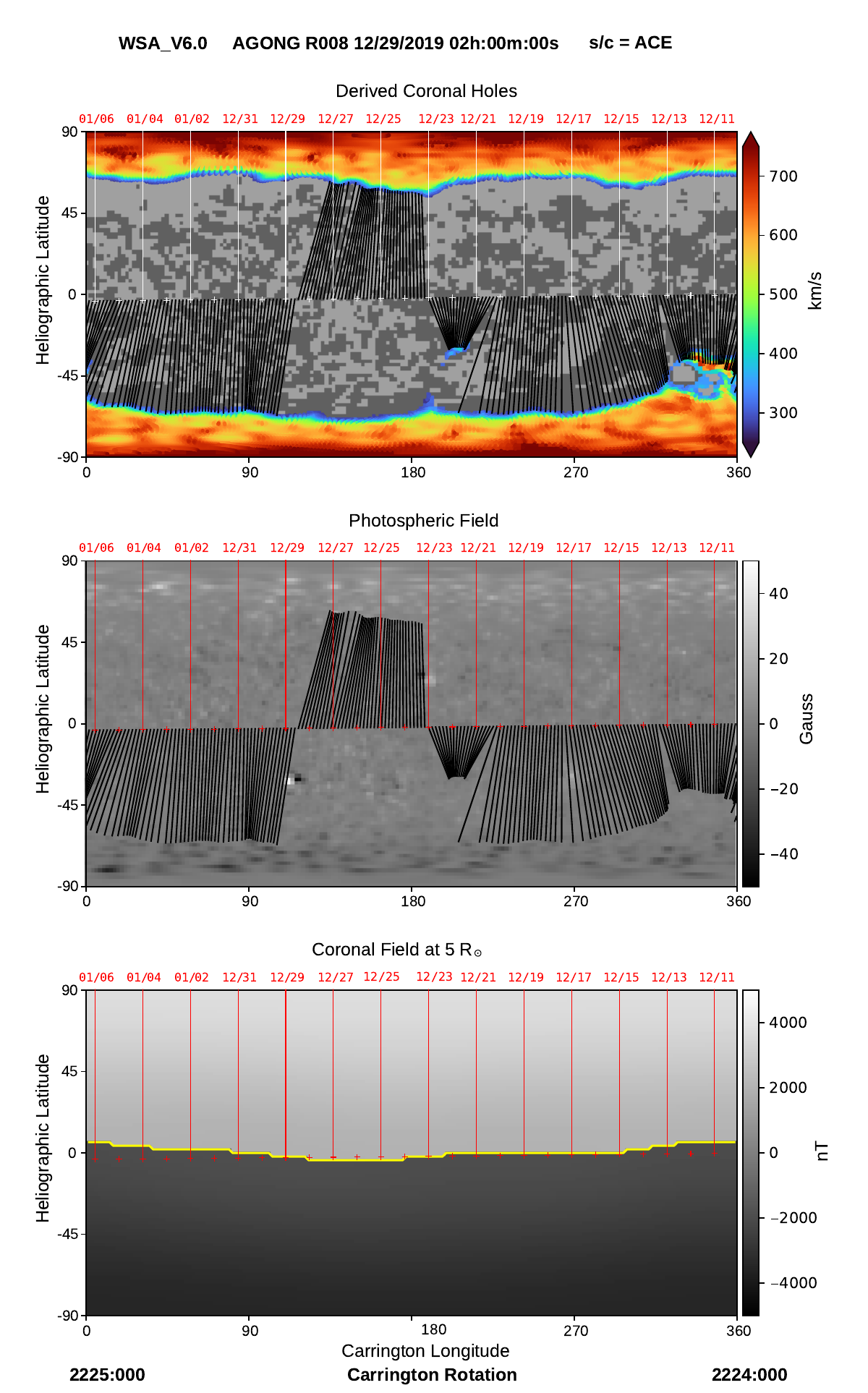}
\caption{ADAPT-WSA model output for CR 2225. The model run is for \textit{Wind} (or ACE) event (corresponding to Figures \ref{fig:1} ). The white (top panel) and red (middle and bottom panel) tick-marks or vertical lines represent the back projection of the \textit{Wind} satellite at 5 $R_\odot$. The top panel shows a WSA-derived open field at 1 $R_\odot$ with model-derived solar wind speed in color scale. Black lines indicate the magnetic connectivity between the projection of the observing satellite location at 5 $R_\odot$ and the solar wind source region at 1 $R_\odot$. The field polarity at the photosphere is indicated by the light/dark (positive/negative) grey contours in the same upper panel. The photospheric field can be seen in the middle panel. This panel's black lines again show the connectivity between the ACE, and the photosphere. The bottom panel shows the WSA-derived coronal field at 5 $R_\odot$. The yellow contour marks the model-derived heliospheric current sheet, where the overall coronal field changes sign.   \label{fig:3}}

\end{center}
\end{figure}

\begin{figure}
\begin{center}
\includegraphics[scale=0.6]{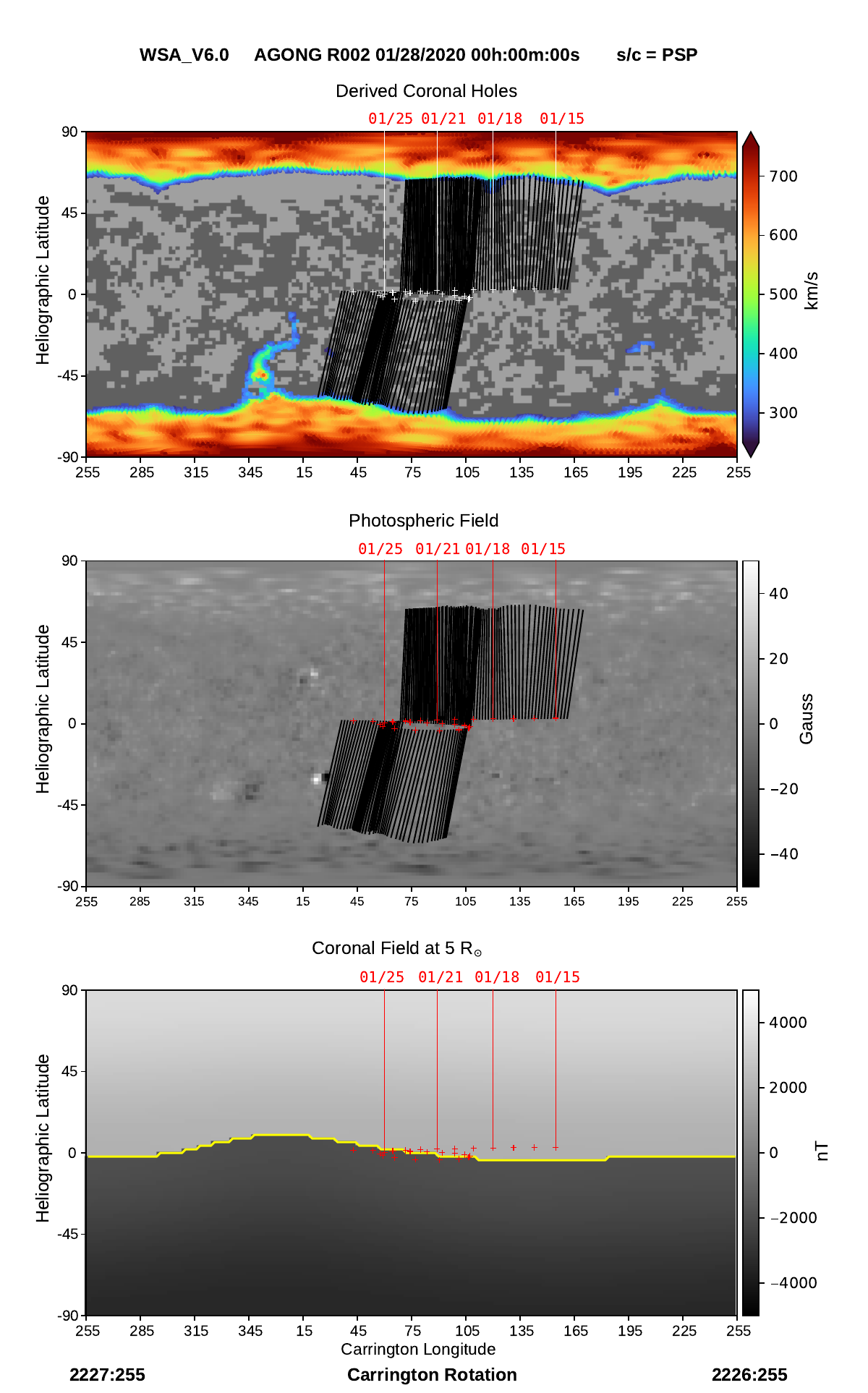}
\caption{Similar to figure \ref{fig:3}, ADAPT-WSA model output for CR 2226. The model run is for PSP event, respectively (corresponding to Figures \ref{fig:2}). The white (top panel) and red (middle and bottom panel) tick-marks or vertical lines represent the back projection of the PSP satellite at 5 $R_\odot$. The top panel shows a WSA-derived open field at 1 $R_\odot$ with model-derived solar wind speed in color scale. Black lines indicate the magnetic connectivity between the projection of the observing satellite location at 5 $R_\odot$ and the solar wind source region at 1 $R_\odot$. The field polarity at the photosphere is indicated by the light/dark (positive/negative) grey contours in the same upper panel. The photospheric field can be seen in the middle panel. This panel's black lines again show the connectivity between the PSP, and the photosphere. The bottom panel shows the WSA-derived coronal field at 5 $R_\odot$. The yellow contour marks the model-derived heliospheric current sheet, where the overall coronal field changes sign.   \label{fig:4}}

\end{center}
\end{figure}

Figures \ref{fig:3} and \ref{fig:4} show the ADAPT-WSA model output of the coronal magnetic field and sources of the \textit{Wind} and PSP events discussed in the previous section and shown in Figures \ref{fig:1} and \ref{fig:2}.  The top panel in Figure \ref{fig:3} and \ref{fig:4} show the WSA-derived coronal holes and the spacecraft connectivity (marked by the black lines) between the projection of \textit{Wind}/PSP's location at 5 $R_\odot$ (red/white tickmarks) and open field foot-points at 1 $R_\odot$.   The black lines reveal the model-derived source regions of the solar wind observed at \textit{Wind}/PSP. The dates in red refer to the location of each spacecraft in time over a Carrington rotation. These dates correspond to when the solar wind left the Sun as opposed to when it arrived at the spacecraft. The middle panel shows the spacecraft connectivity to the photospheric (1 $R_\odot$) magnetic field, and the bottom panel shows the coronal magnetic field at 5 $R_\odot$.  The color bar represents model-derived solar wind speed in the same panel. The three panels are shown for the whole Carrington rotation (Figure \ref{fig:3}: CR 2225, Figure \ref{fig:4}: CR 2226). Figure \ref{fig:3} and \ref{fig:4} shows that the two solar wind events discussed in the previous section that were observed on 28-31 Dec 2019 at Wind and 28-29 Jan 2020 at PSP which originated from a large global helmet streamer connected with the boundaries of coronal holes. 

We also use WSA model parameters derived for each of the 28 \textit{Wind} events and 4 PSP events to further characterize the solar sources that produce low helium abundance in the solar wind and report them in Supplementary Table S2. The model parameters are calculated for the identified field lines that are the source of each event. It is important to note that the model-derived field lines identified as the sources of these events only relate to a 2D slice that the spacecraft connects to, which is a part of a much larger and complex 3D magnetic field topology. Even still, these model-determined sources help inform us about the sources of the low $A_{He}$ events. The model calculations are best matched for the events' duration instead of an entire Carrington rotation. This comparison helps in the improved identification of sources for the specific time periods of our low $A_{He}$ intervals.

We have verified that the sources of almost all the events are coronal hole boundaries as the average minimum angular separation between the field lines identified as the sources of these events and the nearest coronal hole boundary is less than $4^{\circ}$ (see supplementary Table S2). These coronal hole boundaries are generally linked with large-scale quiescent streamers \citep{Higginson2017} which suggests that these events originate from the coronal streamers. Additionally, almost all of these events occur within 10° of the heliospheric current sheet (HCS), which is formed by coronal bipolar (helmet) streamers \citep{Wang2000}. Many of these events are associated with magnetic fields that have moderate to high expansion factors, which act as proxies for solar wind speed. This suggests that the sources of these events are located at or near coronal hole boundaries. Although one might typically expect higher expansion factors at coronal hole boundaries, the expansion factor is directly dependent on the photospheric field strength at the field line footpoint. Most of these events are linked to quiet Sun magnetic fields. Interestingly, the sources of these events are independent of the observation point in the heliosphere, whether observed by PSP or \textit{Wind}. The role of these source regions in the context of low helium abundance events is discussed in the upcoming sections.

\subsection{Comparison of Fe/O and $A_{He}$ }

\begin{figure}
\begin{center}
\plotone{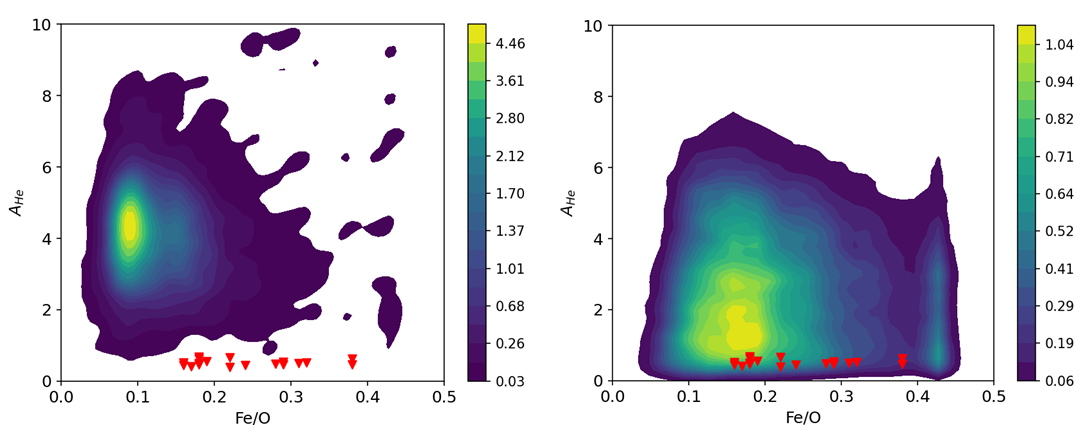}
\caption{The Fe/O vs $A_{He}$ density distributions from ACE SWICS data in fast wind ($>$600 km/s, left) and slow wind ($<$450km/s, right). The red patches show our events' average $A_{He}$ and Fe/O values.    \label{fig:5}}

\end{center}
\end{figure}

In the previous subsection, we found that these events originated from streamers, which can provide plasma from the coronal hole, cusp and legs \citep{Suess2009}. In order to understand the contribution of the coronal hole plasma, we compare the density distribution of Fe/O vs $A_{He}$ for the fast-solar wind events (left panel of Figure \ref{fig:5}) with that during the slow wind events (right panel of Figure \ref{fig:5}). For this purpose, data from the ACE satellite from 1998 to 2020 are used. Because we select 48 hours long intervals, we consider propagation effects between ACE and \textit{Wind} to be minimal. The right panel of Figure \ref{fig:5} is pertinent for the low $A_{He}$ events. We include the fast wind in the left panel for comparison. Figure \ref{fig:5} indicates that the Fe/O ratio of the slow wind events are predominantly located at the higher values of Fe/O. The red patches in both the panels of Figure \ref{fig:5} are $A_{He}$ and Fe/O values averaged for the entire duration of the low $A_{He}$ events considered in the present work. Interestingly, these patches are outside the observed distribution of the fast wind but inside the distribution corresponding to the slow wind. Further, these events significantly deviate from the distribution of fast wind, suggesting a minimal contribution from coronal hole plasma. We also compared the events separately using SWICS 1.1 and SWICS 2.0. There was a shift in the absolute values, but this does not impact our inferences.

\subsection{Signatures of the periodic reconnections}

\begin{figure}
\centering
\gridline{\fig{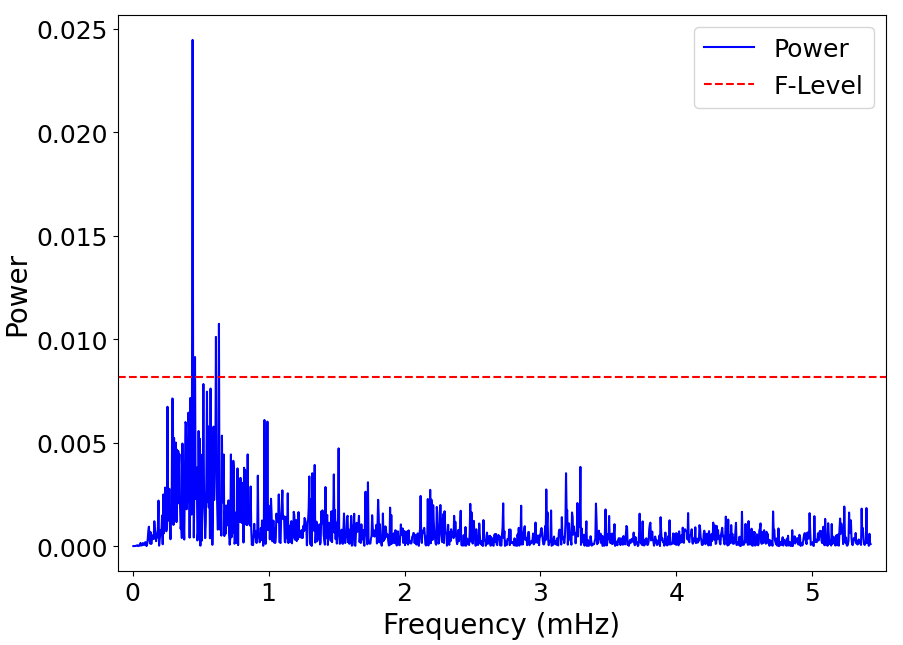}{0.45\textwidth}{(a)}
          \fig{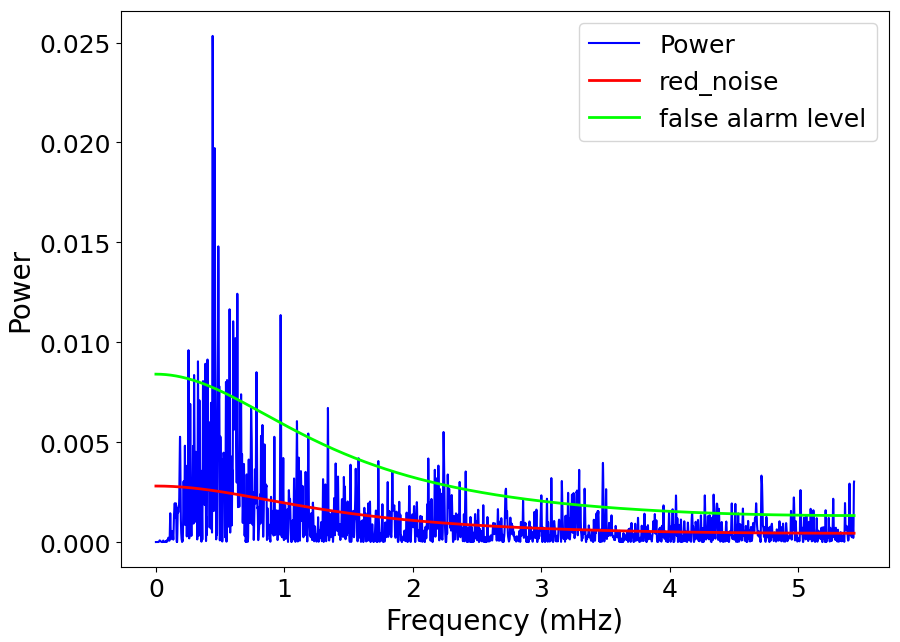}{0.45\textwidth}{(b)}
          }
\caption{The FFT with Fisher's false alarm test (F-level, left panel) and Lomb-Scargle periodogram analysis with red noise and chi-95\% significance level (right panel) for the event shown in Figure \ref{fig:1}. It can be seen in both panels that the frequency of 0.44 mHz (period ~2270 sec) shows the highest power.    \label{fig:6}}
\end{figure}

\cite{Suess2009} suggested that plasma blobs are released through the cusp of the streamers. The sharply pinched magnetic field confines the plasma, and this plasma can be released easily by small pressure pulses. In this section, we have explored the presence of these periodic pressure pulses caused by Alfv\'{e}n waves. We have performed the fast Fourier transform (FFT) and the Lomb-Scargle periodogram on $A_{He}$ time series. Figure \ref{fig:6} shows the frequencies present in the event (number 21) present in Figures \ref{fig:1}. 

We also conducted the Fisher test and red noise levels to check the significance of the periods during these events. Figure \ref{fig:6} shows the FFT with Fisher's false alarm test (F-level, left panel) and Lomb-Scargle periodogram analysis with red noise and chi-95\% significance level (right panel). The frequency of 0.44 mHz (period $\sim$2270 sec) has a high power (above the false alarm and Fisher’s F- level), suggesting the presence of a clear signature of characteristic Alfv\'{e}nic waves. The interchange reconnection model proposed by \cite{Lynch2014} has a characteristic timescale of approximately 2,000 seconds, corresponding to a frequency of about 0.5 mHz, which is comparable to the global Alfvén frequency. This provides indirect evidence to the proposition that episodic release of low $A_{He}$ parcels from the streamer cusps are triggered by Alfv\'{e}n waves. Periodicities around 0.44 mHz were also observed by \cite{Gershkovich2023} in various solar wind compositions. Similar analyses were performed for all \textit{Wind} and PSP events, and almost all events showed similar periods (see supplementary Table S3). The top 5 significant periods are listed in the table.   

\subsection{Low helium abundance in Ulysses observations}

\begin{figure}
\centering
\gridline{\fig{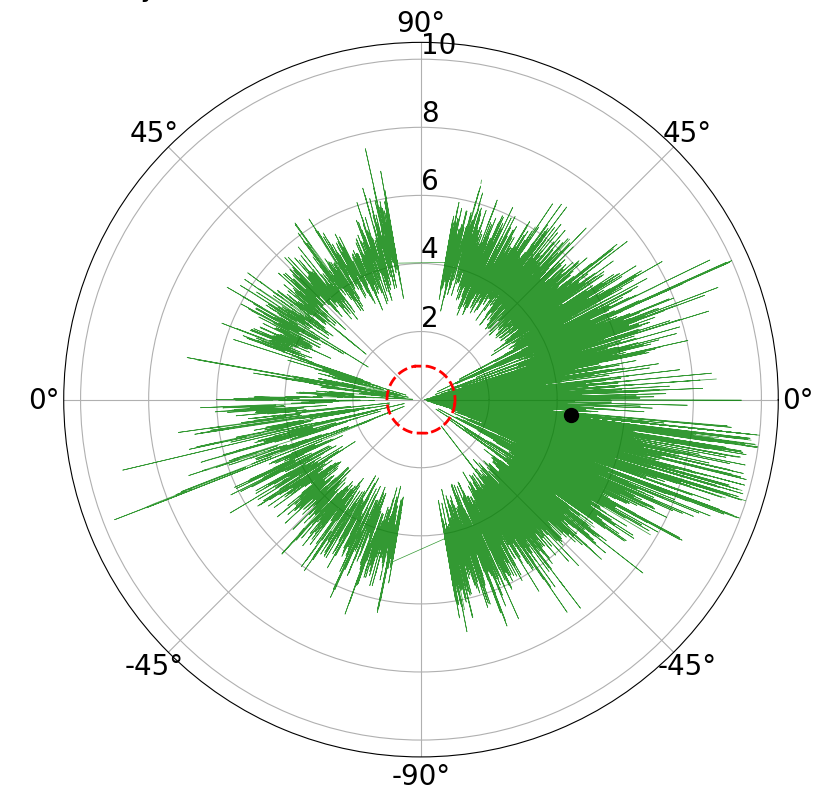}{0.3\textwidth}{(a)}
          \fig{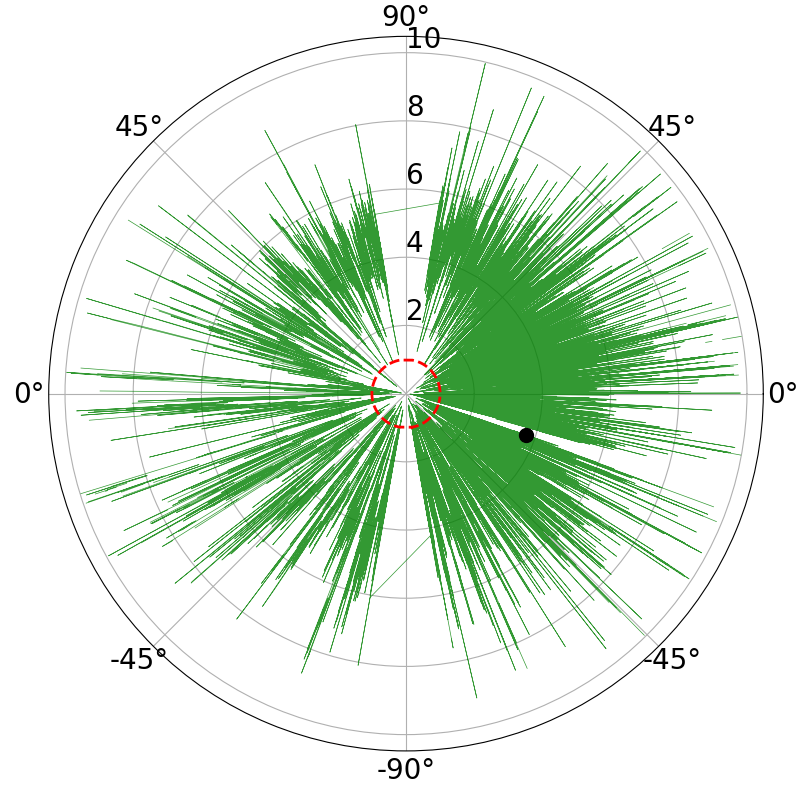}{0.3\textwidth}{(b)}
          \fig{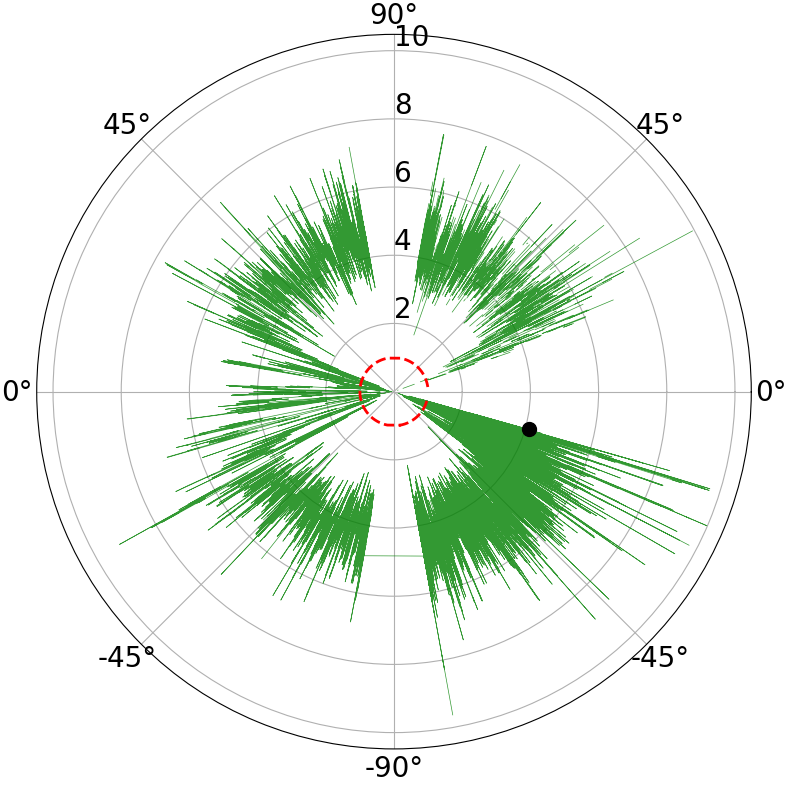}{0.3\textwidth}{(c)}
          }
\caption{Helium abundance variation with heliographic latitudes. The dashed red circle shows $A_{He}$=1\%. The black dot shows the start time and the orbit progress clockwise. The Ulysses first (1992-1998), second (1999-2004) and third (2005-2009) orbit observations are shown. The first and third orbit shows low helium abundance events near the equatorial plane, whereas the 2nd orbit shows the events spread over all the heliographic latitudes.  \label{fig:7}}
\end{figure}

To further characterise the origin of this class of events, we have used Ulysses data to examine the heliographic variations of the helium abundance. Figure \ref{fig:7} shows the $A_{He}$ variation with helio-graphic latitudes. The red circle represents the $A_{He}$=1\%. The first (left panel) and third (right panel) orbit represent the solar minima, whereas the second (middle panel) one shows the solar maxima. It can be observed from Figure \ref{fig:7} that the $A_{He}<$1\% events are present near the equatorial plane during solar minima (Orbit 1 and 3), and these events are distributed towards higher heliolatitudes during the solar maxima (Orbit 2).  

\section{Discussion}
The low helium abundance events are important as they can provide critical information regarding the sources of solar wind and possibly clues on the acceleration mechanism. These events are observed throughout the heliosphere. Coronal hole boundaries are source regions for the slow solar wind \citep{Schwadron2005, Abb16}. The slow solar wind usually has a lower $A_{He}$ than the fast wind and is highly variable  \citep{Kasper2007, Sanchez-Diaz2016}. The analysis presented in the previous section demonstrates that the sources of very low helium events are associated with streamers. A detailed discussion regarding the association of the low $A_{He}$ events with the streamers can also be found in \cite{Suess2009}. \cite{Borrini1981} also showed that low helium abundances are observed near the HCS. Interestingly, our work based on 28 events from \textit{Wind} and 4 from PSP reveal low $A_{He}$ events near the heliospheric current sheet. In addition, almost all \textit{Wind} events characteristically show low alpha density, slow speed, negligible alpha and proton speed difference, and similar temperatures. Therefore, we infer that all the events originated from similar sources in the corona. The source backtracing using the WSA model supports this inference. PSP observations also show similar properties. Similar observations are also reported in \cite{Suess2009}. In their work, \cite{Suess2009} suggested that three potential locations associated with streamers can produce low helium abundance. These locations are the coronal hole, the streamer core just below the magnetic cusp, and the streamer legs (see, Figure 11 of \citealp{Suess2009}). The plasma from these three sources can be supplied in the following manner. First, there is a probability that the plasma from the coronal holes adjacent to streamers can enter the streamer region via Kelvin-Helmholtz (KH) instability \citep{Suess2009}. Second, the streamer core region, which is located just below the cusp, can contribute plasma via reconnection with open field lines. Third, the streamer legs can provide plasma to the streamers through open magnetic field lines. This description is consistent with three-fluid models of the slow solar wind in corona streamers that find gravitational settling of helium (as well as other heavy ions) in the core of streamers reducing their relative abundance in the streamer stalk, and outflow of these heavy ions at the streamer legs, due to the Coulomb friction with slow solar wind stream electrons and protons \citep{Ofm04, OfmKra2010, Abb16}. 

To understand the contribution of coronal hole plasma, we compared the Fe/O vs $A_{He}$ density distribution for fast and slow solar wind events using ACE satellite data from 1998 to 2020 (Figure \ref{fig:5}). It can be seen from Figure \ref{fig:5} that these events significantly deviate from the distribution of fast wind, suggesting a minimal contribution from coronal hole plasma. These events show higher Fe/O ratio, indicating substantial influence of the FIP effect. This also suggests that the plasma comes from longer loops, resulting in a higher FIP processing \citep{Laming2015}. However, as suggested by \cite{Laming2019}, the FIP effect alone cannot explain the low helium abundances; therefore, we explore other possibilities as well. 

The next possibility is that the low $A_{He}$ plasma comes from the streamers' cores or the streamer's legs. \cite{Suess2009} also showed a good correlation between O/H and He/H and argued that O/H is reduced in the core of the streamers compared to the legs. Therefore, they suggested that streamer cores could be potential sources of low $A_{He}$ winds. The low $A_{He}$ events were proposed as transient events. Further, depleted helium abundance is mostly observed towards one edge of the HCS. The streamer core region located just below the cusp can release plasma from a specific side, leading to a depletion of $A_{He}$ on that side of the HCS. 

\begin{figure}
\begin{center}
\plotone{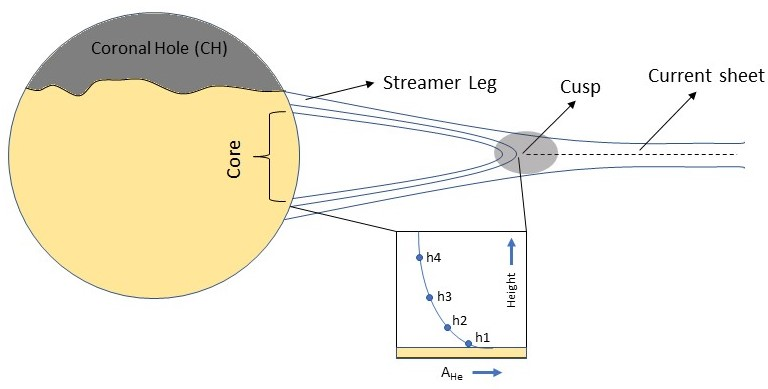}
\caption{The schematic structure of the streamers. The (ideal) gravitationally settled helium abundance profile is shown. The reduction in helium abundance in the solar wind is attributed to the release of plasma from higher altitudes, e.g., the solar wind coming from h4 will have less $A_{He}$ as compared to wind from h1. The zoomed version is rotated by $90^o$.}  \label{fig:8}

\end{center}
\end{figure}

\cite{Suess2009} suggested that plasma blobs are released through the cusp of the streamers. The sharply pinched magnetic field confines the plasma, and this plasma can be released easily by small pressure pulses. These small perturbations cause an episodic release of low $A_{He}$ plasma. The streamer cusp, probably pinching off by reconnection and destabilized by Alfv\'{e}n waves, can release the plasma with low $A_{He}$. The resonant period for such waves (i.e., the travel time from one footpoint to the other) is on the order of $10^3$ seconds (\cite{Gershkovich2023}, coronal loop length/Alfv\'{e}n speed $\sim$ $10^{11}$ cm/$10^8$ cm/s). This period is evident in solar wind signatures and can be seen in Figure \ref{fig:6} and in supplementary Table S3.

Another important point is that the quiescent streamer cores are the only structures stable and quiescent enough to allow gravitational settling, which can lead to a very low $A_{He}$ as well as of other heavy ions in the solar wind as demonstrated by the three-fluid models \citep[e.g.,][]{Ofm04, OfmKra2010, Ofm15, Abb19}. Figure \ref{fig:8} shows three potential sources of solar wind similar to \cite{Suess2009}. The three sources are the streamer leg, coronal hole, and cusp. The core of the streamer can be regarded as the ideal location for gravitational settling to occur. The (ideal) profile of gravitationally settled helium abundance is shown in the square box, which is rotated by $90^o$. Following this in Figure \ref{fig:8}, if we consider gravitational stratification in the streamers, the solar wind coming from height h4 will have lower $A_{He}$ as compared to wind from height h1. So, the solar wind released from the streamer cusps can show a very low helium abundance. Note, in the above discussion, the interplanetary modulation causing the changes in $A_{He}$ \citep{Yogesh2023} is ignored because the low $A_{He}$ events are observed by PSP near the Sun as well as at 1 AU by \textit{Wind}.

The low $A_{He}$ events are also observed by Ulysses. The $A_{He}$ variation during three different Ulysses orbits (Figure \ref{fig:7}) represent different phases of the solar cycle. These distributions are similar to the distribution of the streamers during the solar maxima and minima \citep{Owens2014}. During solar minimum, there are longer duration events with lower $A_{He}$ values compared to solar maximum (not shown in the paper). This intriguing feature can be attributed to the lifetime variability of streamers during the solar cycle \citep{Owens2014}. During solar maximum, the streamers are distributed towards the higher heliolatitudes and are short-lived compared to the long and stable streamers during solar minima. Furthermore, the dominant presence of solar wind originating from streamers typically exhibits lower $A_{He}$ values. So, the Ulysses observations also suggest that the low $A_{He}$ events originate from streamers. This variation of $A_{He}$ in streamers can also explain the solar cycle variation in $A_{He}$, as suggested by \cite{Kasper2007}, where they proposed that the dominance of streamers as the source of $A_{He}$ during solar minima could result in lower $A_{He}$ values compared to solar maxima.

Finally, observing low $A_{He}$ events in the extended solar wind requires specific conditions. The present investigation explains the sources of these low $A_{He}$ events. However, to understand the quantitative aspects, detailed modeling of the streamer cusp regions and the altitudes from which the solar wind originates is still required. This aspect is beyond the scope of this paper.

\section{Conclusions}
The very low helium abundance events ($A_{He} < 1\%$) are a unique feature of the slow solar wind and are observed throughout the heliosphere. These events are generally characterized by very low solar wind speed and negligible differential streaming between the alphas and protons in the case of \textit{Wind} events. In contrast, events observed by PSP show speed that is smaller compared to other times but not negligible. The ADAPT-WSA model analysis indicates that these low $A_{He}$ solar wind parcels originate from quiet Sun coronal helmet streamers. This was also supported using the Ulysses observations. The Ulysses observations showed that these events are distributed along the heliolatitude during the solar maxima, whereas they are distributed near the equator during the solar minima. These distributions are similar to how streamers are distributed during the solar maxima and minima \citep{Owens2014}. In other words, these Ulysses observations are also consistent with the inference that low $A_{He}$ events originate in quiet Sun coronal hole streamers.

It has been proposed in the past that coronal streamers can release plasma through interaction with coronal hole boundaries, streamer legs and streamer core. Here, it has been demonstrated using compositional proxies that the coronal hole plasma is not released from streamer legs. Instead, the streamer cusps situated above the core of the streamer act as a source region. Based on the frequencies observed in $A_{He}$, this release of low $A_{He}$ parcels from the streamer cusps are likely triggered by Alfv\'{e}n waves. The sharply pinched magnetic field confines the plasma, and this plasma can be released easily through magnetic reconnection triggered by small perturbations caused by Alfv\'{e}n waves. The signatures of these waves are also observed in these events. The streamer cores are the only structures stable and quiescent enough to allow gravitational settling, which can lead to a very low $A_{He}$ in the solar wind. This low $A_{He}$ plasma is released from the tops of streamers. The reported low helium abundance is consistent with multi-fluid models of streamers,  demonstrating the gravitational settling of helium ions in quiescent streamer cores and associated depletion of helium in the slow solar wind.

\section{Acknowledgement} 

We thank Prof. J. Martin Laming and Dr. Samantha Wallace for their constructive input and suggestions for this work.  We also thank the WSA model development team at NASA/GSFC for providing model runs for our event list. We thank the PIs of ACE, \textit{Wind} and Ulysses spacecraft. Parker Solar Probe was designed, built, and is now operated by the Johns Hopkins Applied Physics Laboratory as part of NASA's Living with a Star (LWS) program (contract NNN06AA01C). Support from the LWS management and technical team has played a critical role in the success of the Parker Solar Probe mission. We thank NASA/GSFC for providing the facilities to carry out this work. Yogesh also thank SCOSTEP for the opportunity as a Visiting Scholar, under which this work was accomplished. This work is supported by the Department of Space, Government of India. Yogesh and L.O. acknowledge support by NSF grant AGS-2300961, L.O. acknowledges support by NASA LWS grant 80NSSC20K0648. NG is supported by NASA’s LWS program and the STEREO project. PM acknowledges the support from NASA HGIO grant 80NSSC23K0419. 

\bibliography{Low_ahe}{}
\bibliographystyle{aasjournal}

\end{document}